\documentclass[aps,prl,longbibliography,
,groupedaddress, twocolumn,notitlepage]{revtex4-1}

\usepackage{amsmath,amssymb}
\usepackage{natbib}
\usepackage{graphicx,subfigure}
\usepackage{hyperref}
\usepackage[normalem]{ulem}

\newcommand{\be}{\begin{equation}}
\newcommand{\ee}{\end{equation}}
\newcommand{\bse}{\begin{subequations}}
\newcommand{\ese}{\end{subequations}}
\newcommand{\ba}{\begin{eqnarray}}
\newcommand{\ea}{\end{eqnarray}}
\newcommand{\bea}{\begin{eqnarray}}
\newcommand{\eea}{\end{eqnarray}}

\usepackage{color}
\usepackage[normalem]{ulem}  

\begin{document}


\title{Quantum thermodynamics of holographic quenches and bounds on the growth of entanglement from the QNEC}



\author{Tanay Kibe, Ayan Mukhopadhyay and Pratik Roy}
\email[]{ayan@physics.iitm.ac.in, tanayk@smail.iitm.ac.in, pratik@physics.iitm.ac.in}
\affiliation{Center for Quantum Information Theory of Matter and Spacetime, and Center for Strings, Gravitation and Cosmology, Department of Physics, Indian Institute of Technology Madras, Chennai 600036, India}

\date{\today}

\begin{abstract}
The quantum null energy condition (QNEC) is a lower bound on the energy-momentum tensor in terms of the variation of the entanglement entropy of a sub-region along a null direction. To gain insights into quantum thermodynamics of many-body systems, we study if the QNEC restricts irreversible entropy production in quenches driven by energy-momentum inflow from an infinite memoryless bath in two-dimensional holographic theories. We find that an increase in both entropy and temperature, as implied by the Clausius inequality of classical thermodynamics, are necessary but not sufficient to not violate QNEC in quenches leading to transitions between thermal states with momentum which are dual to Banados-Teitelboim-Zanelli geometries. For an arbitrary initial state, we can determine the lower and upper bounds on the increase of entropy (temperature) for a fixed increase in temperature (entropy). Our results provide explicit instances of quantum lower and upper bounds on irreversible entropy production whose existence has been established in literature. We also find monotonic behavior of the non-saturation of the QNEC with time after a quench, and analytically determine their asymptotic values. Our study shows that the entanglement entropy of an interval of length $l$ always thermalizes in time $l/2$ with an exponent $3/2$. Furthermore, we determine the coefficient of initial quadratic growth of entanglement analytically for any $l$, and show that the slope of the asymptotic ballistic growth of entanglement for a semi-infinite interval is twice the difference of the entropy densities of the final and initial states. We determine explicit upper and lower bounds on these rates of growth of entanglement.
\end{abstract}

\maketitle

Quantum thermodynamics has established various instances where thermodynamics can be generalized even to finite-dimensional quantum systems interacting with a bath by accounting for entanglement and measures of accessible quantum information \cite{PhysRevLett.115.070503,Guryanova_2016,PhysRevE.93.022126,Yunger_Halpern_2016,Goold_2016,Gour_2018,RevModPhys.91.025001}. For instance, it has been shown that the one-shot work cost of creating a state and the extractable work from it are bounded by the hypothesis-testing relative entropy between the state and the thermal equilibrium \cite{PhysRevE.93.022126}. This discipline has found applications in understanding (bio-)chemical reactions, and also in the study of quantum engines.

Although limited progress has been achieved in the applications of quantum thermodynamics to many-body systems, there have been independent developments of interest. One such example is the formulation of the Quantum Null Energy Condition (QNEC) \cite{Bousso_2016} which sets lower bounds on the expectation value of null components of the energy-momentum tensor in terms of null variations of the entanglement entropy of subregions whose boundary contains the point of observation. QNEC has been proven for free quantum field theories (QFTs) \cite{Bousso_2016_2,Malik:2019dpg}, holographic QFTs \cite{Koeller_2016}, two-dimensional (2D) conformal field theories (CFTs) \cite{Balakrishnan:2017bjg}, and also for general Poincar\'e-invariant QFTs using half-sided modular inclusion properties of operator algebras \cite{Ceyhan:2018zfg}. In a 2D CFT with central charge $c$, the strictest form of QNEC is \cite{Wall:2011kb,Koeller_2016,Balakrishnan:2017bjg}
\begin{equation}\label{Eq:QNEC}
\mathcal{Q}_\pm \equiv 2\pi \langle t_{\pm\pm}\rangle - S_{\rm ent}'' - \frac{6}{c}{S_{\rm ent}'}^2 \geq 0,
\end{equation}
where $ t_{\pm\pm}$ are the two non-vanishing null components of the energy-momentum tensor, and the derivatives are obtained from infinitesimal variations of the entanglement entropy $S_{\rm ent}$ of any interval ending at the point of observation under shifts along the $+$ (right) and $-$ (left) pointing null directions respectively. Recently it has been pointed out that QNEC can follow from positivity conditions on variations of the relative entropy under null shape deformations \cite{Leichenauer:2018obf,Lashkari:2018nsl,Moosa:2020jwt} (see also \cite{Ceyhan:2018zfg}) and such positivity conditions also hold for sandwiched Renyi divergences. A pertinent question is therefore, whether QNEC and its possible generalizations impose criteria which go beyond classical thermodynamics, such as quantum generalizations of the Clausius inequality discussed in the literature \cite{PhysRevLett.105.170402,PhysRevLett.107.140404,Plastina_2014,March_2016,PhysRevLett.107.140404,Van_2021,RevModPhys.93.035008}.


When any system with finite energy interacts with a memoryless infinitely large energy bath, its entropy can only increase monotonically. In holographic QFTs, this feature is reproduced via the monotonic growth of the area of the apparent and event horizons, and eventual thermalization following a quench \cite{Bardeen:1973gs,Balasubramanian:2010ce,Hubeny:2010ry,Chesler:2013lia}.  

In this work, we consider fast quenches that lead to transitions between thermal states carrying momentum in 2D holographic systems, and establish that the QNEC implies more than the mere rise of both the temperature and the thermodynamic entropy. For a fixed increase in the entropy (temperature), the increase in temperature (entropy) has to be bounded from both above and below so that the QNEC is unviolated after quench.  Our results thus provide explicit instances of the upper and lower bounds on irreversible entropy production in quantum many-body systems whose existence has been established using tools of quantum information theory \cite{PhysRevLett.105.170402,PhysRevLett.107.140404,Plastina_2014,March_2016,PhysRevLett.107.140404,Van_2021,RevModPhys.93.035008}. Furthermore, we extend previous results \cite{Calabrese:2004eu,Calabrese:2005in,Calabrese:2009qy,Hubeny:2013hz,Liu:2013iza,Liu:2013qca,Rangamani:2015agy,Leichenauer:2015xra,Calabrese:2016xau} on the growth and thermalization of entanglement entropy in 2D CFTs, and establish bounds on the rates of growth of entanglement which can be validated in numerical simulations and experiments. 

Our holographic results apply when the central charge $c$ of the CFT is large and it has a sparse spectrum (implying strong coupling). Nevertheless, we argue that our results are relevant for understanding entropy production from quenches faster than any microscopic time scale in a generic many-body system.
\section{Holographic quenches} 
A two-dimensional strongly coupled holographic CFT with a large central charge can be described by a three-dimensional Einstein gravity coupled to a few fields and with a negative cosmological constant \cite{Aharony:1999ti}. The central charge of the dual CFT is $c = 3L/(2G)$ \cite{Brown:1986nw,Henningson:1998gx,Balasubramanian:1999re}, where $G$ is Newton's gravitational constant and $L$ is related to the cosmological constant $\Lambda$ via $\Lambda =-1/L^2$. Any (time-dependent) state in the CFT corresponds to a regular solution of the gravitational theory.

Quenches leading to fast transitions between thermal states at time $t=0$ can be described by dual metrics of the form (see also \cite{Sfetsos:1994xa})
\begin{eqnarray}\label{Eq:metric}
{\rm d}s^2 &=& 2 {\rm d}r {\rm d}t -\left(\frac{r^2}{L^2} - 2 m(t)L^2\right) {\rm d}t^2 + 2 j(t) L^2{\rm d}t {\rm d}x  \nonumber\\ &&+\frac{r^2}{L^2} {\rm d}x^2, 
\end{eqnarray}
with
\begin{align}
 m(t) &= \theta(-t)({\mu_+^{i}}^{2} + {\mu_-^{i}}^{2}) + \theta(t) ({\mu_+^{f}}^{2} + {\mu_-^{f}}^{2}),\\
 j(t) &= \theta(-t)({\mu_+^{i}}^{2} - {\mu_-^{i}}^{2}) + \theta(t) ({\mu_+^{f}}^{2} - {\mu_-^{f}}^{2})
\end{align}
where $\mu^{i,f}_{\pm}$ are related to the temperature ($T^{i,f}$) and entropy density ($s^{i,f}$) of the initial and final thermal states respectively. Explicitly,
\begin{equation}\label{Eq:T-s}
T^{i,f} = \frac{2}{\pi} \frac{\mu_+^{i,f} \mu_-^{i,f}}{\mu_+^{i,f}+ \mu_-^{i,f}}, \quad s^{i,f} =\frac{c}{6} \left(\mu_+^{i,f}+ \mu_-^{i,f}\right).
\end{equation}
These can be obtained from the thermodynamics of the Banados-Teitelboim-Zanelli (BTZ) black branes \cite{Banados:1992wn,Banados:1992gq} dual to the initial and final states. The coordinates $t$ and $x$ are shared by the dual field theory which lives at the boundary $r = \infty$ of the emergent radial direction.
This geometry is supported by a bulk stress tensor $T_{MN}$ that is traceless and locally conserved in the metric \eqref{Eq:metric} with non-vanishing components
\begin{equation}\label{Eq:Tbulk}
T_{tt} = \frac{q(t) L^2}{r}  + \frac{p(t) j(t)L^6}{r^3},\quad T_{tx} = \frac{p(t)L^2}{r},
\end{equation}
where
\begin{eqnarray}\label{Eq:q-p}
8\pi G q(t) &=& \delta(t)\left({\mu_+^{f}}^{2} - {\mu_+^{i}}^{2} +{\mu_-^{f}}^{2} - {\mu_-^{i}}^{2} \right), \nonumber\\ 
8\pi G p(t) &=& \delta(t)\left({\mu_+^{f}}^{2} - {\mu_+^{i}}^{2} -{\mu_-^{f}}^{2} + {\mu_-^{i}}^{2} \right).
\end{eqnarray}
We find that the QNEC inequalities \eqref{Eq:QNEC} imply that the bulk matter satisfies the classical null energy condition.

Holographic renormalization \cite{Henningson:1998gx,Balasubramanian:1999re} of the on-shell gravitational action for the metric \eqref{Eq:metric} provides the expectation value of the energy-momentum tensor of the dual state (living in flat Minkowski metric):
\begin{equation}\label{Eq:thol}
\langle t_{\pm \pm} \rangle =\frac{c}{12 \pi}\left(\theta(-t) {\mu_{\pm}^{i}}^2  + \theta(t){\mu_{\pm}^{f}}^2 \right),\quad \langle t_{+-} \rangle =0.
\end{equation}
The vanishing of $\langle t_{+-} \rangle$ implies tracelessness. Gravitational constraints \eqref{Eq:q-p} also imply the Ward identity  $\partial_\mu \langle t^{\mu\nu} \rangle = f^\nu$, where $f_\nu =L (q(t,x), p(t,x))$ is the energy-momentum injection from the infinite bath into the CFT.

Finally, we note that we have been agnostic about the matter content of the bulk theory while describing the dual geometries. The explicit form \eqref{Eq:q-p} of the bulk energy-momentum tensor, which is localized on the ingoing null shell, is simply necessitated by the Israel junction conditions. Our results therefore do not depend on the specific details of the dual CFTs.

\subsection{The cut and glue method}
\begin{figure}[t]
\includegraphics[scale=.25]{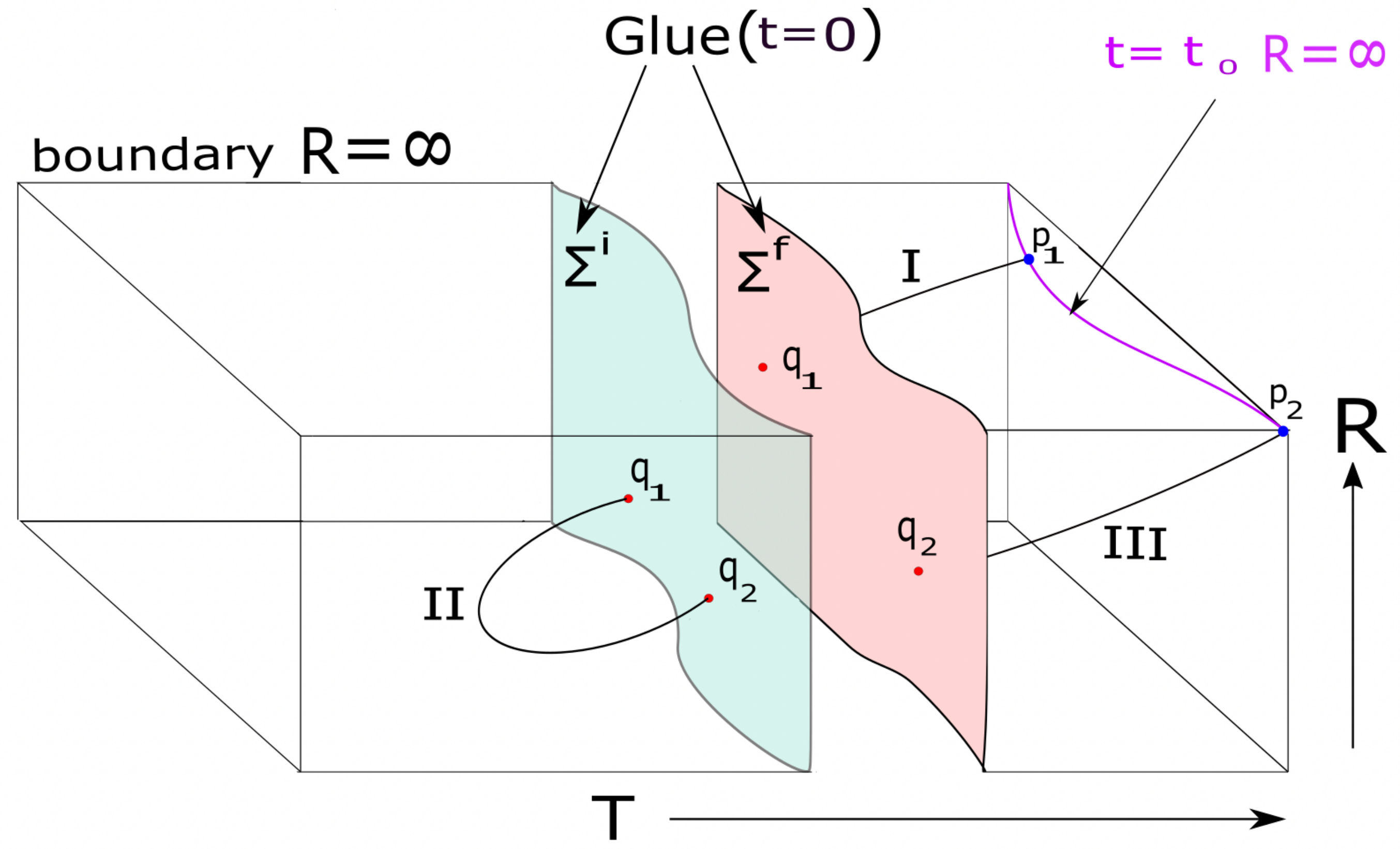}
\caption{Schematic representation of the cut and glue method -- the left and right halves represent the separate Poincar\'{e} patches to which the pre-quench and post-quench spacetimes map to. The gluing hypersurfaces $\Sigma^{i,f}$ are the images of $t=0$ in the respective geometries. Points on $\Sigma^{i,f}$ carrying the same physical coordinate labels $x$ and $r$ are identified. The geodesic ending at the boundary is cut into three arcs.
}\label{Fig:Glue}
\end{figure}

For our analytic computations, we use the result that the geometry \eqref{Eq:metric} describing a fast transition between two BTZ black branes at $t=0$ can be \textit{uniformized}, i.e. converted to the Poincar\'{e} patch metric (with $m(t) = j(t) =0$ in \eqref{Eq:metric}),
corresponding to the vacuum, with two \textit{separate} diffeomorphisms for $t<0$ and $t>0$ (see Supplemental Material for details). These uniformization maps result in two Poincar\'{e} patches bounded by the hypersurfaces ($\Sigma^{i,f} (x,r)$) that are the respective images of the hypersurface $t=0$. These hypersurfaces are glued by identifying the points on each with the same values of the \textit{physical} coordinates $x$ and $r$. See Fig.~\ref{Fig:Glue} for an illustration.

\section{Entanglement growth}
The entanglement entropy of a spacelike interval with end points ${p}_1 = (x_1, t_1)$ and  ${p}_2 = (x_2, t_2)$ in any arbitrary state of the holographic CFT can be obtained from the proper length $L_{\rm geo}$ of the geodesic in the dual bulk geometry which is anchored to the points $p_1$ and $p_2$ at the regulated boundary $r=L^2 /\epsilon$, and is given by $S_{\rm ent} = (c/6)(L_{\rm geo}/L)$ \cite{Ryu:2006bv,Hubeny:2007xt}. Here, $\epsilon^{-1}$ is interpreted as an ultraviolet (UV) energy cutoff in the dual theory. Since the geodesic length can be readily computed in the Poincar\'{e} patch metric given just the endpoints \footnote{This computation can be readily done by the geodetic distance formula between two end points $(T_1,X_1,R_1)$ and $(T_2,X_2,R_2)$ in the Poincar\'{e} patch metric. This geodetic distance $L_{geo}$ is simply $L\ln (\xi + \sqrt{\xi^2 -1})$ where
$\xi = (Z_1^2 + Z_2^2 - (T_1 +Z_1 - T_2 -Z_2)^2 + (X_1 - X_2)^2)/2 Z_1 Z_2$ with $Z_{1,2} = L^2/R_{1,2}$.}, we can compute the entanglement entropy of a spatial interval of length $l$ in any BTZ state employing the uniformization map. A simple computation  yields 
\begin{equation}
    S_{\rm ent} = \frac{c}{6}\ln\left(\frac{\sinh(\mu_+l)\sinh(\mu_-l)}{\mu_+\mu_-l^2}\right) + \frac{c}{3}\ln \left(\frac{l}{\epsilon}\right),
\end{equation}
where the last term is the well-known vacuum contribution that depends on the UV regulator (see \cite{Holzhey:1994we,Calabrese:2004eu,Cadoni:2010ztg}).

Via the cut and glue method we can readily compute the evolution of entanglement entropy for intervals at the boundary of geometries describing fast quenches between two BTZ spacetimes. Before the quench, the geodesic is located entirely in a single Poincar\'{e} patch described by the uniformization map for $t<0$. After the quench, the geodesic anchored to the boundary (at points $p_{1,2}$ in Fig.~\ref{Fig:Glue}) of the final Poincar\'{e} patch goes back in time and intersects the gluing hypersurface $\Sigma^{f}$, at two points ($q_{1,2}$ in Fig.~\ref{Fig:Glue}) until a limiting time where the two intersection points merge, after which the entanglement entropy for the chosen interval thermalizes. The hypersurface $\Sigma^f$ thus cuts the geodesic into three arcs, two of which (labelled I and III in Fig.~\ref{Fig:Glue}) are in the final Poincar\'{e} patch and join $p_{1,2}$ to $q_{1,2}$ on $\Sigma^f$, and a third arc (labelled II in Fig.~\ref{Fig:Glue}) joins $q_{1,2}$ on $\Sigma^i$ in the initial Poincar\'{e} patch. The length of each of the three geodetic arcs can be computed by the Poincar\'{e} patch distance formula since the endpoints are known explicitly. Variations of the entanglement entropy under null deformations of any of the endpoints can similarly be computed. For more details see Supplemental Material.

Our explicit computations confirm that the entanglement entropy has three stages of evolution \cite{Hubeny:2013hz,Liu:2013iza,Liu:2013qca} for a transition between two BTZ spacetimes. In the first stage, the entanglement entropy of an interval of length $l$ grows quadratically from its pre-quench values as $\sim D_s t^2$ with
\begin{eqnarray}\label{Eq:Ds}
    D_s &=& \frac{c}{6}\Bigg(\Delta m+ 2\left(\mu_+^f \coth(\mu_+^f l)-\mu_+^i\coth(\mu_+^i l)\right)\nonumber\\&&\left(\mu_-^f \coth(\mu_-^f l)-\mu_-^i\coth(\mu_-^i l)\right)\Bigg),
\end{eqnarray}
where $\Delta m ={\mu_+^f}^2 +{\mu_-^f}^2 -{\mu_+^i}^2 -{\mu_-^i}^2$. The above reproduces the known result for the vacuum to thermal non-rotating BTZ transition \cite{Hubeny:2013hz,Liu:2013qca}. This initial quadratic growth has also been observed in quantum lattice systems \cite{Unanyan:2010,Unanyan:2014}. In the intermediate regime, the entanglement grows quasi-linearly. For a semi-infinite interval, the asymptotic growth is exactly linear, i.e. $S_{\rm ent} = v_s t$ with 
\begin{equation}\label{Eq:vs}
    v_s = 2 (s^f - s^i)
\end{equation}
where $s^{i,f}$ are the initial (final) entropy densities given by \eqref{Eq:T-s}. This is consistent with the \textit{tsunami hypothesis} \cite{Liu:2013iza,Calabrese:2016xau} (see also \cite{Asplund:2015eha,vonKeyserlingk:2017dyr,Mezei:2016wfz}) that the entanglement \textit{spreads} with a tsunami velocity, which is the speed of light in 2D CFTs \cite{Calabrese:2009qy,Calabrese:2016xau}, from both ends of the interval so that sub-intervals of total length $2t$ should become completely entangled with the rest of the quenched system. If an interval of large length can be approximated by a thermal density matrix (see \cite{Calabrese:2009qy,Hubeny:2013gta,Kudler-Flam:2021rpr,Kudler-Flam:2021alo}), then the result \eqref{Eq:vs} follows because the change in the entanglement at late time should be the product of the length $2t$ times the difference in the thermodynamic entropy densities between final and initial states (see \cite{Mandal:2014wfa,Erdmenger:2017gdk} for other contexts). This light-cone like spreading of entanglement has been observed analytically in CFTs using replica methods \cite{Calabrese:2005in}, numerically in quantum lattice systems \cite{Kim:2013,PhysRevA.89.031602,PhysRevLett.111.260401,PhysRevLett.111.207202,Chiara_2006}, and experimentally in ultra-cold atomic gases \cite{Cheneau2012,Langen2013} and ion traps \cite{Jurcevic:2014,Richerme:2014,Bonnes:2014}. Our general result \eqref{Eq:vs} can thus be validated both numerically and experimentally.

We are also able to prove that the entanglement entropy $S^{ent}(t)$ for any interval of length $l$ saturates sharply to the thermal value $S_{\rm th}$ at the so called \textit{horizon time} $t =l/2$ and also seen in lattice simulations as $ S_{\rm th} - S^{ent}(t)\sim (l/2 -t)^{3/2}$ as $t\rightarrow l/2$ for arbitrary fast quenches. This readily follows from the analytic result that the final intersection point between $\Sigma^f$ and the geodesic glued to the endpoints of the interval $0 \leq x\leq l$ at the boundary in the post quench geometry occurs at $t=l/2$, and is given by the point on $\Sigma_f$ parametrized by
\begin{equation}\label{Eq:ExtremeIntersection}
    r_* = L^2(\mu_+^f\coth(\mu_+^f l)+\mu_-^f\coth(\mu_-^f l)), \, x_* = l/2,
\end{equation}
The horizon time and the saturation exponent $3/2$ were found earlier in holographic systems only for the transition from the vacuum to a non-rotating thermal state \cite{Hubeny:2013hz,Liu:2013iza,Liu:2013qca}. However, this feature can be shown to be valid analytically for a class of quenches in generic 2D CFTs \cite{Calabrese:2005in,Calabrese:2009qy,Calabrese:2016xau} and is also seen in experiments \cite{Cheneau2012}. It will be interesting to also reproduce our general results from tensor network approaches building on \cite{Hartman:2013qma}.


\section{The QNEC Criterion}
It can be readily seen that the momentum carrying thermal states dual to BTZ geometries saturate the QNEC inequalities \eqref{Eq:QNEC} for any length $l$ of the entangling interval \cite{Ecker:2019ocp}. Therefore, these inequalities should be saturated before the quench. However, after the quench time ($t=0$), we find that the QNEC inequalities \eqref{Eq:QNEC} can be violated.

We find that the QNEC inequalities \eqref{Eq:QNEC} imply the strictest bounds when applied for the semi-infinite interval (see Supplemental Material for dependence of $\mathcal{Q}_\pm (t)$ on the length $l$ of the entangling interval).
Translation symmetry further implies that it is sufficient to consider intervals $x\geq 0$ with $\mathcal{Q}_+$ ($\mathcal{Q}_-$) involving null variations of the endpoint at the spatial origin towards right (left) respectively. Applying the cut and glue method for the semi-infinite interval, we see that demanding $\mathcal{Q}_\pm \geq 0$ at $t=0$ implies (with $\Delta = 2(\mu_+^f-\mu_+^i)(2\mu_+^f+\mu_+^i)$):
\begin{align}   \label{Eq:mufallowed}
    & \frac{1}{3}\left(\sqrt{\Delta+3\mu_-^i(3\mu_-^i +2\mu_+^i -2\mu_+^f)}+
    \mu_+^f -\mu_+^i \right)        \\
     & \leq \mu_-^f  \leq \sqrt{\Delta +\mu_-^i(\mu_-^i+2\mu_+^f - 2 \mu_+^i)} - \mu_+^f +\mu_+^i. \nonumber			
\end{align}
For the initial vacuum state ($\mu_\pm ^i =0$), the above inequalities simply impose that $\mu_+^f = \mu_-^f$, i.e. the final state should have zero momentum. One can analytically show that for the latter case $\mathcal{Q}_\pm = 0$ for all time in the case of the semi-infinite interval (see Supplemental Material). It is quite interesting that although the thermalization of the the entanglement occurs at the tsunami speed (of light), QNEC saturation persists throughout the quench.

\begin{figure}[h]
\setlength{\abovecaptionskip}{17.5 pt plus 4pt minus 2pt}
\mbox{
\subfigure{\includegraphics[scale=0.46]{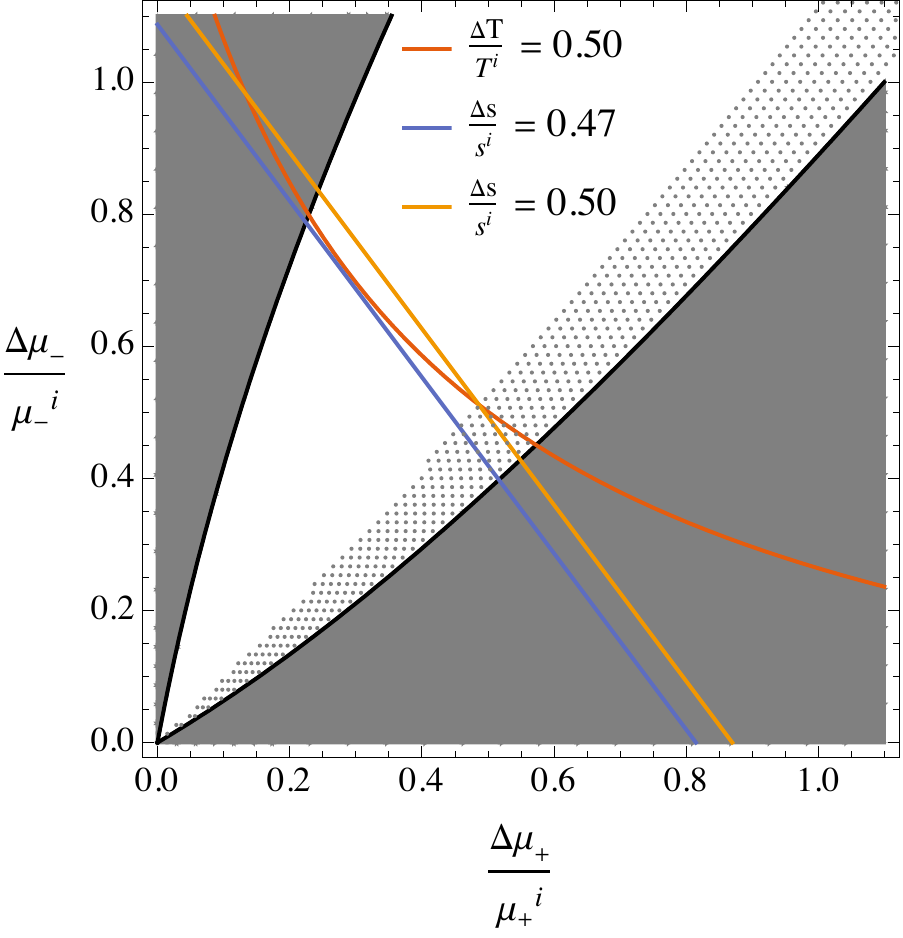}}\,\,
\subfigure{\includegraphics[scale=.46]{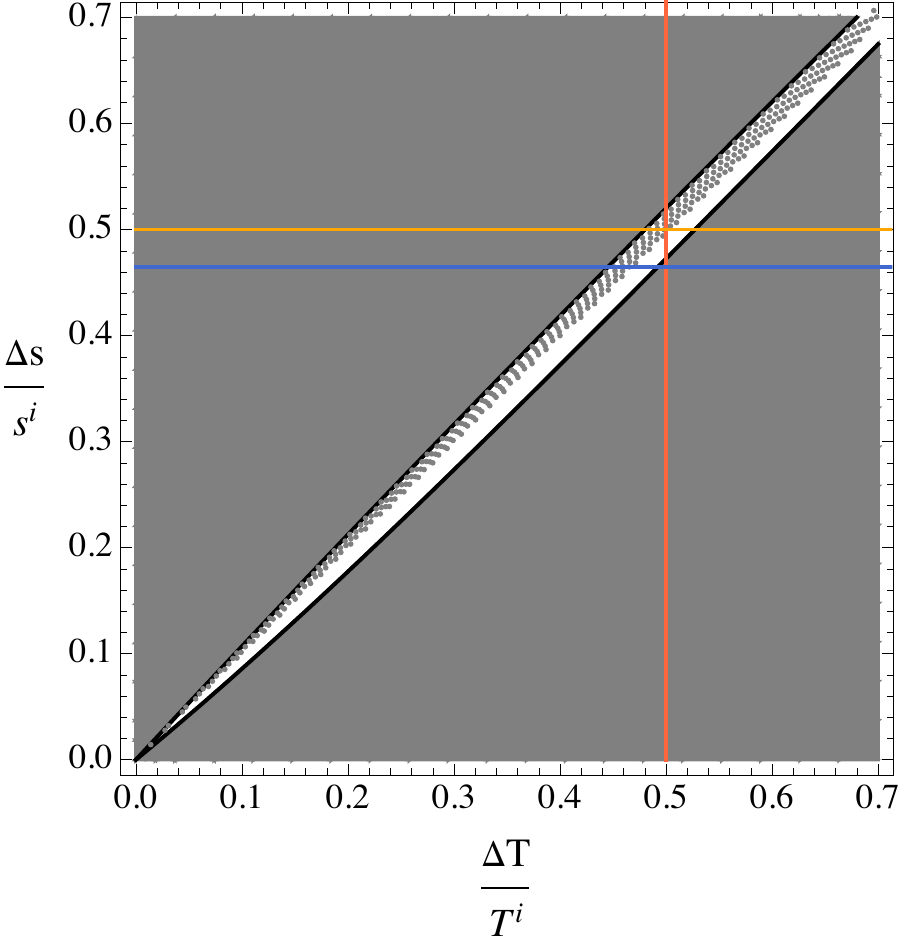}}
}
\vspace{-0.7 cm}
\caption{Left: The possible relative changes $\frac{\mu_{\pm}^{f}-\mu_{\pm}^{i}}{\mu_{\pm}^{i}}$ for $\mu_+^i = 1, \, \mu_-^i = 0.75$ are shown in white. The black lines are given by the inequality \eqref{Eq:mufallowed} required by $\mathcal{Q}_\pm \geq 0$ at $t=0$ and the grey dotted region is disallowed by examining $\mathcal{Q}_-$ for $t>0$. The contours show that for a fixed change in temperature (red), the change in entropy is bounded from above and below (yellow and blue). Right: The same allowed region (in white) is shown in terms of the relative change in temperature and entropy density.}
\label{Fig:QNEC}
\end{figure}

When the initial state is not the vacuum, the inequality \eqref{Eq:mufallowed} implies that $\mu_\pm^f \geq\mu_\pm^i $, and therefore $T^f > T^i$ and $s^f > s^i$, i.e. both the temperature and thermodynamic entropy density must not decrease after quench. However, as stated before, we get more. As for instance, with $\mu_+^i = 1, \, \mu_-^i = 0.75$ the final states satisfying \eqref{Eq:mufallowed} lie within the region bounded by the black bold lines shown at the  left in Fig.~\ref{Fig:QNEC} implying stricter bounds than classical thermodynamics. 
When the upper (lower) end of the inequality \eqref{Eq:mufallowed} is satisfied, $\mathcal{Q}_+$ ($\mathcal{Q}_-$) vanishes at $t=0$ for the semi-infinite interval. For $t>0$, although $\mathcal{Q}_+ >0$ is always satisfied when \eqref{Eq:mufallowed} holds,  $\mathcal{Q}_- \geq 0$ is violated for $ t> t_c$ (with $t_c$ depending on initial and final states) thus pushing above the lower bound on $\mu_-^f$ set by \eqref{Eq:mufallowed}, depicted by the upper boundaries of the dotted regions in Fig.\ref{Fig:QNEC} (see Supplemental Material for more details). The final allowed region, shown in white in Fig.~\ref{Fig:QNEC}, implies lower and upper bounds on the increase in  entropy density (temperature) for a fixed increase in temperature (entropy density). The corresponding plot of allowed final states for the initial state $\mu_\pm^i =1$ shown in the Supplemental Material also illustrates these bounds.

Furthermore, we find that as $t\rightarrow \infty$,
\begin{equation}\label{Eq:QNECplateau}
    \mathcal{Q}_- \rightarrow0, \,\,  \mathcal{Q}_+\rightarrow (s^f -s^i)( \mu_+^f -\mu_-^f+\mu_+^i + \mu_-^i ) >0
\end{equation}
for the semi-infinite interval. Interestingly, for allowed transitions, $\mathcal{Q}_+(t)$ and $\mathcal{Q}_-(t)$ for the semi-infinite interval are also monotonically increasing and decreasing functions respectively after quench. Its implications for the relative entropy of the quenched state should be understood following \cite{Casini:2017roe,Lashkari:2018nsl} (see also \cite{Leichenauer:2018obf,Moosa:2020jwt}). 

The bounds shown in Fig.~\ref{Fig:QNEC}  can be understood in terms of irreversible entropy production. For any process the total change in entropy can be decomposed as $\Delta S = \Delta S_{\rm irr} + \Delta S_{\rm rev}$, where $\Delta S_{\rm rev}$ is the entropy change due to reversible heat exchange with a bath. The Clausius inequality implies that $\Delta S_{\rm irr}  \geq 0$. This have been generalized in quantum thermodynamics. Ref. \cite{PhysRevLett.105.170402} provides a lower bound on $\Delta S_{\rm irr}$ in terms of the Bures distance between the out-of-equilibrium state and the final equilibrium state, and an upper bound related to the Bremermann-Bekenstein bound \cite{PhysRevLett.46.623}. These bounds can be equivalently stated in terms of the average irreversible work \cite{PhysRevLett.107.140404,Plastina_2014,March_2016}. Bounds on $\Delta S_{\rm irr}$ have also been seen for an open quantum system coupled to a thermal bath \cite{PhysRevLett.107.140404,Van_2021,RevModPhys.93.035008}. However, such bounds depend on the choice of a distance measure on the space of states, and it is not clear which one places the tightest bound. The fast quenches considered here do not involve any reversible heat exchange, implying $\Delta S = \Delta S_{\rm irr}$. Our results provide an explicit computation of lower and upper bounds on $\Delta S_{\rm irr}$ for a fixed change in temperature in a strongly interacting many-body system. 

The upper and lower bounds on $\Delta s$, the increase in entropy density, for a fixed final temperature and a given initial state, readily bound the speed of the asymptotic ballistic entanglement growth \eqref{Eq:vs} for the semi-infinite interval from both above and below. Furthermore, the coefficient of the initial quadratic growth \eqref{Eq:Ds} is similarly bounded from above and below for any $l$, and both of these bounds increase monotonically with $l$ (see Supplemental Material for plots).

Our results should be valid when the time scale of the quench is smaller than any other scale in the system, and the final and initial temperature scales are both smaller than the microscopic energy scale below which the CFT provides a good description \cite{Calabrese:2016xau, Buchel:2013gba}. Since the strongest bounds on irreversible entropy production and entanglement growth correspond to the semi-infinite interval, our results are insensitive to the microscopic details. Our bounds can thus be verified qualitatively from growth of entanglement by studying fast quenches e.g. in spin-$\frac12$ XX and XXZ chains numerically \cite{Unanyan:2010,Unanyan:2014}, and experimentally in ultra-cold atomic gases \cite{Cheneau2012,Langen2013} and in ion traps \cite{Jurcevic:2014,Richerme:2014,Bonnes:2014}. Going beyond the requirements $c\gg1$ and a sparse spectrum would require investigating higher derivative and quantum corrections in the gravitational description.

\section{Discussion}

Our result establishing lower and upper bounds on irreversible entropy production in holographic CFTs after quenches via the application of QNEC offers a novel perspective on the quantum thermodynamics of many-body systems.

As detailed in the Supplemental Material, our methods allow study of fast quenches between arbitrary \textit{quantum equilibrium states} \cite{Ecker:2019ocp} which saturate the QNEC inequalities \eqref{Eq:QNEC}. These states can be essentially described as Virasoro hair on top of vacuum and thermal states, and are dual to Banados geometries \cite{Banados:1998gg}. Following our methods, an erasure protocol for quantum information encoded in Virasoro hair has been implemented \cite{lumps-paper}. The QNEC inequalities reproduce the Landauer principle \cite{sagawa2017second,Esposito_2010,Reeb_2014} and also demonstrate that certain types of encoding are tolerant against erasures faster than any microscopic timescale \cite{lumps-paper}. A more general study of transitions between quantum equilibrium states should lead to novel consequences for various quantum channels.

\begin{acknowledgments}
It is a pleasure to thank Shira Chapman, Christian Ecker, Daniel Grumiller, Arul Lakshminarayan, Prabha Mandayam, Marios Petropoulos, Giuseppe Policastro and Suhail Ahmad Rather for helpful discussions. We also thank Souvik Banerjee for collaboration during early stages of this work, and Avik Banerjee and Nehal Mittal for collaboration on further investigations to appear in a future publication. The research of TK is supported by the Prime Minister's Research Fellowship (PMRF). AM acknowledges the support of the Ramanujan Fellowship of the Science and Engineering Board of the Department of Science and Technology of India, the new faculty seed grant of IIT Madras and the additional support from the Institute of Eminence scheme of IIT Madras funded by the Ministry of Education of India. 
\end{acknowledgments}

\newpage
\section*{Supplemental Material}
\subsection{Quenched Ba\~nados geometries and the uniformization map}
In the main text we describe transitions between rotating BTZ black branes. These are a special case of transitions between the more general Ba\~nados geometries \cite{Banados:1998gg} defined below. 
Quenches leading to transitions between quantum equilibrium states (dual to Ba\~nados geometries) can be described by dual metrics of the form (see also \cite{Sfetsos:1994xa})
\begin{eqnarray}\label{Eq:metric1}
{\rm d}s^2 &=& 2 {\rm d}r {\rm d}t -\left(\frac{r^2}{L^2} - 2 m(t,x)L^2\right) {\rm d}t^2 + 2 j(t,x) L^2{\rm d}t {\rm d}x  \nonumber\\ &&+\frac{r^2}{L^2} {\rm d}x^2 
\end{eqnarray}
in the ingoing Eddington-Finkelstein gauge. This geometry is supported by a bulk stress tensor $T_{MN}$ that is conserved and traceless in the background metric \eqref{Eq:metric1} and whose non-vanishing components are
\begin{eqnarray}\label{Eq:Tbulk1}
T_{tt} &=& \frac{q(t,x) L^2}{r} + \frac{\partial_x p(t,x) L^4}{r^2} + \frac{p(t,x) j(t,x)L^6}{r^3},\nonumber\\
\,\, T_{tx} &=& \frac{p(t,x)L^2}{r}.
\end{eqnarray}
The gravitational equations (with $\Lambda =-1/L^2$)
\begin{equation}\label{Eq:Eom}
R_{MN} - \frac{1}{2}R G_{MN} - \frac{1}{L^2}G_{MN} = 8\pi G\, T_{MN}
\end{equation}
are satisfied simply by requiring 
\begin{eqnarray}\label{Eq:constraints}
\partial_t m(t,x) -\partial_x j(t,x) &=& 8\pi G q(t,x), \nonumber\\ \partial_t j(t,x) -\partial_x m(t,x)&=& 8\pi G p(t,x).
\end{eqnarray}
In absence of bulk matter, we therefore obtain
\begin{eqnarray}\label{Eq:QEQ}
m(t,x) &=& \mathcal{L}_+ (x^+) + \mathcal{L}_- (x^-), \nonumber\\ j(t,x) &=& \mathcal{L}_+ (x^+) - \mathcal{L}_-(x^-).
\end{eqnarray}
with $x^\pm = t \pm x$, and $\mathcal{L}_\pm$ being arbitrary chiral functions. These solutions are known as Banados geometries \cite{Banados:1998gg} and are related to the vacuum solution (for which $m(t,x) = j(t,x) = 0$) locally by a diffeomorphism that is non-vanishing at the boundary and is dual to a conformal transformation as will be shown later. See \cite{Compere:2015knw,Sheikh-Jabbari:2016unm} for detailed CFT interpretation. The dual CFT states have been called quantum equilibrium states since they saturate the QNEC inequalities \eqref{Eq:QNEC} \cite{Ecker:2019ocp}.

The gravitational constraints \eqref{Eq:constraints} imply that the metric \eqref{Eq:metric} can describe instantaneous transitions between arbitrary quantum equilibrium states at $t = 0$ with the initial ($i$) and final ($f$) chiral functions being $\mathcal{L}_\pm^{i,f}$ so that
\begin{eqnarray}\label{Eq:M-J}
m &=& \theta(-t)(\mathcal{L}^i_+ (x^+) + \mathcal{L}^i_- (x^-)) 
+\theta(t)(\mathcal{L}^f_+ (x^+) + \mathcal{L}^f_- (x^-)), \nonumber\\
j &=& \theta(-t)(\mathcal{L}^i_+ (x^+) - \mathcal{L}^i_- (x^-)) 
+\theta(t)(\mathcal{L}^f_+ (x^+) - \mathcal{L}^f_- (x^-)),
\end{eqnarray}
if we set
\begin{eqnarray}\label{Eq:q-p}
8\pi G q &=& \delta(t)(\mathcal{L}^f_+ (x) - \mathcal{L}^i_+ (x) +\mathcal{L}^f_- (-x) - \mathcal{L}^i_- (-x)), \nonumber\\ 
8\pi G p &=& \delta(t)(\mathcal{L}^f_+ (x) - \mathcal{L}^i_+ (x) -\mathcal{L}^f_- (-x) + \mathcal{L}^i_- (-x)).
\end{eqnarray}

Holographic renormalization \cite{Henningson:1998gx,Balasubramanian:1999re} of the on-shell gravitational action for the metric \eqref{Eq:metric} can be used to extract the expectation value of the energy-momentum tensor of the dual state (living in flat Minkowski metric) and it is (with $c = 3L/2G$):
\begin{equation}\label{Eq:thol}
\langle t_{\pm \pm} \rangle =\frac{c}{24 \pi}( m(t,x) \pm j(t,x)),\quad \langle t_{+-} \rangle =0.
\end{equation}

Any Ba\~nados geometry \eqref{Eq:QEQ} can be uniformized to the Poincar\'{e} patch metric. The uniformization map is defined in terms of boundary lightcone coordinates $X_b^\pm(x^\pm)$ that satisfy
\begin{equation}\label{Eq:Sch}
   {\rm Sch}( X_b^\pm(x^\pm), x^\pm) = - 2 \mathcal{L}_\pm(x^\pm)
\end{equation}
where ${\rm Sch}$ denotes the Schwarzian derivative. The uniformization map, which takes the metric \eqref{Eq:metric} to the Poincar\'{e} patch metric with coordinates $T$, $X$ and $R$ is
\begin{eqnarray}\label{Eq:uniformization}
T &=& \frac{1}{2}\Bigg(X_b^+(x^+) +X_b^-(x^-)\nonumber\\ &&+ \frac{{X_b^+}'(x^+) +{X_b^-}'(x^-)-2 \sqrt{{X_b^+}'(x^+){X_b^-}'(x^-)}}{\frac{r}{L^2}-\frac{{X_b^+}''(x^+)}{2{X_b^+}'(x^+)}-\frac{{X_b^-}''(x^-)}{2{X_b^-}'(x^-)}}\Bigg),\nonumber\\
X &=& \frac{1}{2}\Bigg(X_b^+(x^+) -X_b^-(x^-)+ \frac{{X_b^+}'(x^+) -{X_b^-}'(x^-)}{\frac{r}{L^2}-\frac{{X_b^+}''(x^+)}{2{X_b^+}'(x^+)}-\frac{{X_b^-}''(x^-)}{2{X_b^-}'(x^-)}}\Bigg),\nonumber\\
R &=& L^2\frac{{\frac{r}{L^2 }-\frac{{X_b^+}''(x^+)}{2{X_b^+}'(x^+)}-\frac{{X_b^-}''(x^-)}{2{X_b^-}'(x^-)}}}{\sqrt{{X_b^+}'(x^+){X_b^-}'(x^-)}}.
\end{eqnarray}
It is easy to see that at the boundary $r = \infty$, $X^\pm = T\pm X$ reduces to $X_b^\pm(x^\pm)$ respectively -- so the boundary lightcones are reparametrized following \eqref{Eq:Sch}. The boundary metric, which is identified with the physical metric on which the CFT lives, still remains the flat Minkowski space. The uniformization map transforms $\langle t_{\pm \pm} \rangle$ of the dual quantum equilibrium state to zero in agreement with the conformal anomaly of the dual CFT, provided the central charge is indeed $c = 3L/(2G)$. For BTZ black brane solutions we choose
\begin{equation}\label{Eq:BTZX}
  X_b^\pm(x^\pm) =L \exp(2\mu_\pm x^\pm).
\end{equation}
The uniformization map is valid only outside the outer horizon $r_{out}= L^2(\mu^+ + \mu^-)$ with the latter mapping to the Poincar\'{e} horizon $R=0$. Note that \eqref{Eq:Sch} determines $X_b^\pm(x^\pm)$ only up to an overall $SL(2,R)$ (fractional linear) transformation, which accounts for the local  $SL(2,R)\times SL(2,R)$ isometries of Banados metrics. It is convenient to fix these redundancies by choosing $X_b^\pm$ such that $\infty$ is a fixed point, $X_b^\pm(-\infty) = 0$ and $X_b^\pm(0) = L$.

\subsection{Computing the entanglement entropy and QNEC\\ by the cut and glue method}
Let $x_{1,2}^\pm$ denote the lightcone coordinates of the endpoints of the entangling interval whose entanglement entropy is of interest. In the dual Banados spacetime we place these points at the boundary $r= \infty$. When we map this spacetime to the Poincar\'{e} patch metric via the uniformization map \eqref{Eq:uniformization}, the endpoints get mapped to $(X_{b}^+(x_1^+), X_{b}^-(x_1^-))$ and $(X_{b}^+(x_2^+), X_{b}^-(x_2^-))$ at the boundary $R= \infty$. These endpoints are denoted as $p_1$ and $p_2$ in Fig.~\ref{Fig:Glue}. In what follows, it is useful to redefine the Poincar\'{e} patch radial coordinate as $Z = L^2/R$ so that the boundary is at $Z=0$. Explicitly, the endpoints of the entangling interval at the boundary are at
\begin{align}\label{Eq:Endpoints}
   & T_1 = \frac{1}{2} (X_{b}^+(x_1^+)+X_{b}^-(x_1^-)), X_1 = \frac{1}{2} (X_{b}^+(x_1^+)-X_{b}^-(x_1^-)), \nonumber\\& Z_1 =0,\nonumber\\
   &T_2 = \frac{1}{2} (X_{b}^+(x_2^+)+X_{b}^-(x_2^-)), X_2 = \frac{1}{2} (X_{b}^+(x_2^+)-X_{b}^-(x_2^-)), \nonumber\\ &Z_2 =0.
\end{align}
The equations for the geodesic in the Poincar\'{e} patch metric connecting any two endpoints $(T_1,X_1,Z_1)$ and $(T_2,X_2,Z_2)$ are given by the two equations
\begin{align}\label{Eq:Geodesic}
   & Z + T  - \frac{T_1 + T_2}{2}- \frac{T_2 - T_1}{X_2 - X_1}\left(X - \frac{X_1 +X_2}{2}\right) =0,\nonumber\\
   &\frac{Z^2}{(X_2 - X_1)^2 - (T_2 - T_1)^2}-\frac{(2 X - X_1 - X_2)^2}{4(X_2 - X_1)^2} = \frac{1}{4}.
\end{align}
As shown in Fig.~\ref{Fig:Glue}, when the endpoints lie in the post-quench geometry, the above geodesic is cut into three segments by the cut and glue prescription described in the main text.
The two gluing hypersurfaces, which are simply the respective images of the hypersurface $t=0$ are:
\begin{align}\label{Eq:Sigmas}
    \Sigma^{i,f} = (T^{i,f}(0,x,r),X^{i,f}(0,x,r),R^{i,f}(0,x,r)). 
\end{align}
The intersection points of the geodesic and the gluing hypersurface $\Sigma^f$ parametrized by the physical coordinates $x$ and $r$ can be found by substituting \eqref{Eq:Endpoints} and \eqref{Eq:Sigmas} into the above two equations \eqref{Eq:Geodesic} for the geodesic, and solving for $x$ and $r$. We obtain at most two pairs of solutions giving the two intersection points $q_1(x_1^*, r_1^*)$ and $q_2(x_2^*, r_2^*)$ shown in Fig.~\ref{Fig:Glue}. The image of the intersection points on $\Sigma^i$ is obtained simply by substituting the values of $x_{1,2}^*$ and $r_{1,2}^*$ for the equation describing $\Sigma^i$ in \eqref{Eq:Sigmas}. The geodesic with these endpoints in the pre-quench Poincar\'{e} patch is then completed by the arc following the equations \eqref{Eq:Geodesic} with endpoints chosen to be these intersection points (see Fig.~\ref{Fig:Glue}).  To compute the length of the arcs we simply need the endpoints explicitly and employ the invariant distance formula mentioned in the main text.
    
For BTZ geometries, the explicit form of $X_b^\pm$ given by \eqref{Eq:BTZX} allows for explicit analytic solutions of the intersection points at early time and near thermalization time ($l/2$) generically, and for all times when the final state is non-rotating.  When the boundary interval is at $t = l/2$ the geodesic intersects $\Sigma^f$ tangentially and lies entirely in the post-quench Poincar\'{e} patch otherwise and the two intersection points merge into one as given by
\begin{equation}\label{Eq:ExtremeIntersection}
    r_* = L^2(\mu_+^f\coth(\mu_+^f l)+\mu_-^f\coth(\mu_-^f l)), \, x_* = l/2.
\end{equation}
For $t> l/2$, the geodesic does not intersect the gluing hypersurface. When $t\sim 0$, the intersection points can be computed in power series in $t$. 

It is obvious that the above computation can be repeated after displacing one of the two boundary endpoints to obtain $\mathcal{Q}_\pm$.

We should be able to use this method also to study retarded correlation functions in quenched geometries following \cite{Banerjee:2016ray} -- in this case one needs to analytically continue the bulk fields across the gluing surface. 

\subsection{Explicit results for $S_{\rm ent}$ and $\mathcal{Q}_\pm$ for quenches leading to a non-rotating BTZ state}

Since the intersection points $q_{1,2}$ can be computed for all time $t$ for a final non-rotating BTZ state, we can obtain the time-dependent entanglement entropy for any quench from an arbitrary rotating BTZ (with parameters $\mu_\pm^i$) to final non-rotating BTZ state (with $\mu_\pm^f = \mu$). The final expression is however very cumbersome and simplifies only in the limit when the entangling region has large length $l$. Explicitly, 
\begin{eqnarray}
    S_{\rm ent} &=& \frac{c}{6}\ln \Big( 
    \frac{4\mu^2\cosh^2 (2\mu t)- (\mu_+^i-\mu_-^i)^2\sinh^2 (2\mu t)}{4 \mu^{2}\cosh (2\mu t)^{\frac{\mu_+^i + \mu_-^i}{\mu}}}\Big)\nonumber\\
    && + \frac{c}{6}\ln\left(\frac{\sinh(\mu_+^i l) \sinh(\mu_-^i l)}{\epsilon^{2} \mu_+^i \mu_-^i}\right) + \cdots,
\end{eqnarray}
with $\cdots$ denoting terms which vanish as $l\rightarrow\infty$. The above agrees with the rates of entanglement growth given by \eqref{Eq:Ds} (with $l\rightarrow \infty$) and also with \eqref{Eq:vs} for $\mu_\pm^f =\mu$. It also reproduces the known result for vacuum to non-rotating BTZ transition \cite{Hubeny:2013hz,Liu:2013qca}. 

Similarly one can obtain $\mathcal{Q}_\pm(t)$ analytically for any entangling length $l$ for all time explicitly when the final state is thermal and non-rotating. Note that unlike the entanglement entropy, $\mathcal{Q}_\pm(t)$ is discontinuous at $t=0$ due to the discontinuity of the derivatives of $S_{\rm ent}$ at the quenching time. Explicitly for $t>0$ and  $l\rightarrow\infty$ (i.e. for the semi-infinite interval $x\geq 0$), 
\begin{eqnarray}\label{Eq:QNECnonrot}
    \mathcal{Q}_+ &=& \frac{c}{24}\,\frac{e^{4\mu t}(\mu_+^i + \mu_-^i){\rm sech}^2(2\mu t)}{2\mu - (\mu_+^i - \mu_-^i)\tanh(2\mu t)}
    \Big(2 \mu(2\mu + \mu_-^i - 3\mu_+^i) \nonumber\\&&+( \mu_+^i - \mu_-^i)(2\mu + \mu_+^i +\mu_-^i)\tanh(2\mu t)\Big),\nonumber\\
    \mathcal{Q}_- &=& \frac{c}{24}\,\frac{e^{-4\mu t}(\mu_+^i + \mu_-^i){\rm sech}^2(2\mu t)}{2\mu - (\mu_+^i - \mu_-^i)\tanh(2\mu t))}
     \Big(2 \mu(2\mu + \mu_+^i - 3\mu_-^i) \nonumber\\&&+( \mu_+^i - \mu_-^i)(2\mu + \mu_+^i +\mu_-^i)\tanh(2\mu t)\Big).
\end{eqnarray}
Requiring $\mathcal{Q}_\pm$ to be non-negative at $t=0$ is equivalent to \eqref{Eq:mufallowed} for $\mu_-^f = \mu_+^f = \mu$. When the initial state is the vacuum, $\mathcal{Q}_\pm$ vanishes for all times for the semi-infinite interval as evident from \eqref{Eq:QNECnonrot}, and as claimed in the main text. For an arbitrary initial state, \eqref{Eq:QNECnonrot} implies that $\mathcal{Q}_+$ plateaus at $t\rightarrow\infty$ to a finite value while $\mathcal{Q}_-$ vanishes in agreement with \eqref{Eq:QNECplateau}.
 
\begin{figure}[h]
\includegraphics[scale=.6]{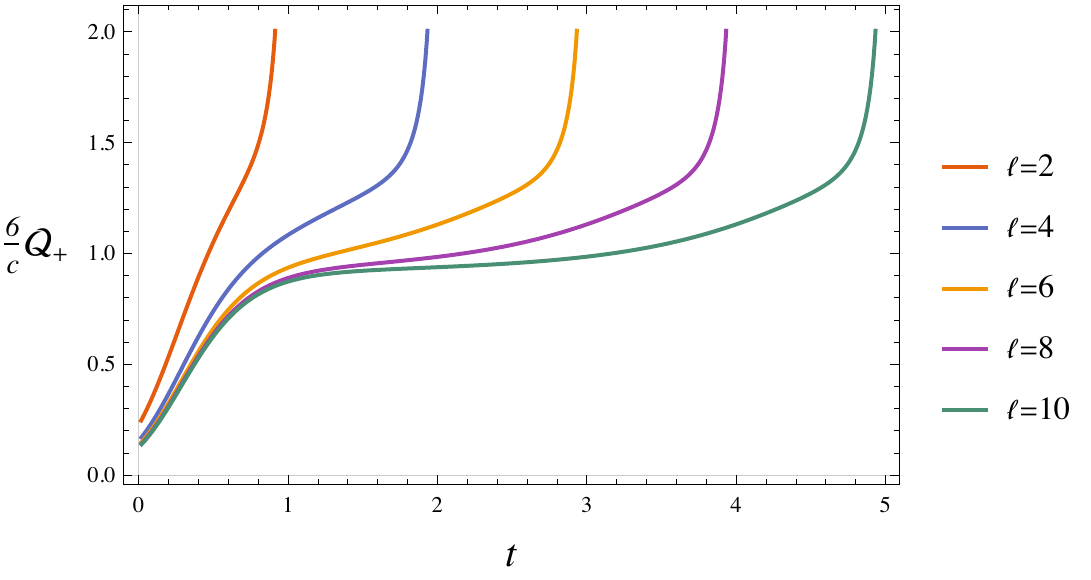}\,
\includegraphics[scale=.6]{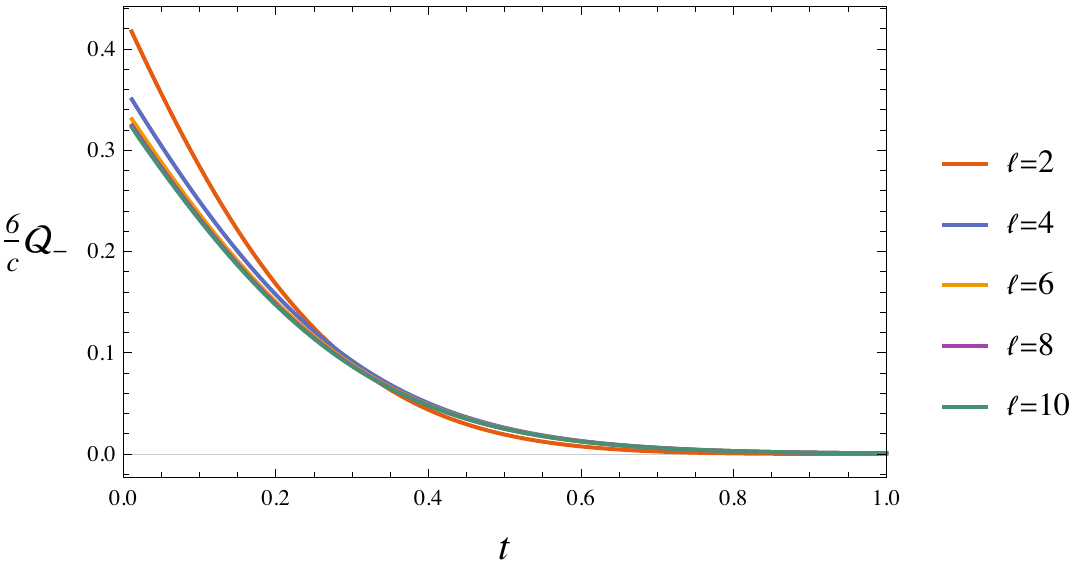}
\caption{$\mathcal{Q}_+$ (top) is plotted for $0< t < l/2$ and $\mathcal{Q}_-$ (bottom) is plotted for $ 0< t < {\rm min}(1, l/2)$ for various lengths $l$ of the entangling interval and the instantaneous quench from $\mu_+^i = 0.5, \mu_-^i = 0.2$ to the non-rotating final state with $\mu_\pm^f = 1$. It is clear that the strictest inequality is imposed in the limit $l\rightarrow\infty$. Note $\mathcal{Q}_\pm$ vanishes for $t<0$ and $t>l/2$.}\label{Fig:ldependence}
\end{figure}

For finite $l$ and arbitrary initial and final states, $\mathcal{Q}_+$ diverges as $(l/2 -t)^{-1/2}$ and $\mathcal{Q}_-$ vanishes as $t\rightarrow l/2$; and both vanish for $t> l/2$ due to thermalization of the entanglement entropy. The coefficient $-1/2$ of the divergence of $\mathcal{Q}_+$ is simply a consequence of the saturation exponent $3/2$ of the entanglement entropy at $t =l/2$ mentioned earlier (and is the result of the double derivative in \eqref{Eq:QNEC}). Plots of $\mathcal{Q}_\pm$ for a representative transition to a final thermal non-rotating state are shown in Fig.~\ref{Fig:ldependence} as a function of time ($t> 0$) for different values of $l$ of the entangling interval $0\leq x \leq l$. It is clear from the plots that the inequality \eqref{Eq:QNEC} gets stricter uniformly for $t> 0$ as $l\rightarrow\infty$. This also holds for arbitrary transitions. We reproduce features of the numerical results for vacuum to non-rotating BTZ transitions for finite $l$ reported in \cite{Ecker:2019ocp}.

\subsection{Computing $\mathcal{Q}_\pm$ for the semi-infinite interval and determining allowed quenches}

For arbitrary quenches between rotating thermal states, the QNEC inequality \eqref{Eq:QNEC} is the strictest for the semi-infinite interval as described above. For a generic instantaneous quench between thermal rotating states, we can compute $\mathcal{Q}_\pm$ for this semi-infinite interval analytically just after the quench i.e. when $t > 0$ and is small, and also when $t$ is large, as in both cases we can obtain analytic expressions for the intersection points between the geodesic and the gluing hypersurface. At $t=0$, $\mathcal{Q}_\pm$ jumps discontinuously from zero to the finite values  given by 
\begin{eqnarray}\label{Eq:Qnecatt01}
	\mathcal{Q}_+(t = 0^+) &=&\frac{c}{24}\big(3({\mu_+^f}^2-{\mu_+^i}^2)\\\nonumber&&+ (\mu_-^f -\mu_-^i)(2\mu_+^i - 2\mu_+^f -\mu_-^f-\mu_-^i)\big),
	\end{eqnarray}
and
	\begin{eqnarray}\label{Eq:Qnecatt02}
	\mathcal{Q}_-(t = 0^+)	&=&	\frac{c}{24}\big(3({\mu_-^f}^2-{\mu_-^i}^2)\\\nonumber&&+ (\mu_+^f -\mu_+^i)(2\mu_-^i - 2\mu_-^f -\mu_+^f-\mu_+^i)\big).
\end{eqnarray}
These are in agreement with the result for $\mathcal{Q}_\pm(t)$ given by \eqref{Eq:QNECnonrot} when evaluated at $t=0$ and with $\mu_\pm^f =\mu$. Note that  $\mathcal{Q}_\pm(t=0^+)$ are related by simultaneous interchanges of $\mu_+^{i}$ with $\mu_-^{i}$ and $\mu_+^{f}$ with $\mu_-^{f}$, i.e. the left and right moving temperatures. (This is not true for $t>0$.) Requiring $\mathcal{Q}_\pm(t=0^+) \geq 0$ and using \eqref{Eq:Qnecatt01} and \eqref{Eq:Qnecatt02} we obtain the inequalities \eqref{Eq:mufallowed}. We will call the upper bound on $\mu_-^f$ given by \eqref{Eq:mufallowed} as $\mu_-^{u}$ and the lower bound as $\mu_-^{l}$. Both $\mu_-^{u}$ and $\mu_-^l$ are determined by $\mu_\pm^i$ and $\mu_+^f$. We study $\mathcal{Q}_\pm(t)$ by numerical determinations of the intersection points at intermediate times for generic transitions.

\begin{figure}[h]
\includegraphics[scale=.35]{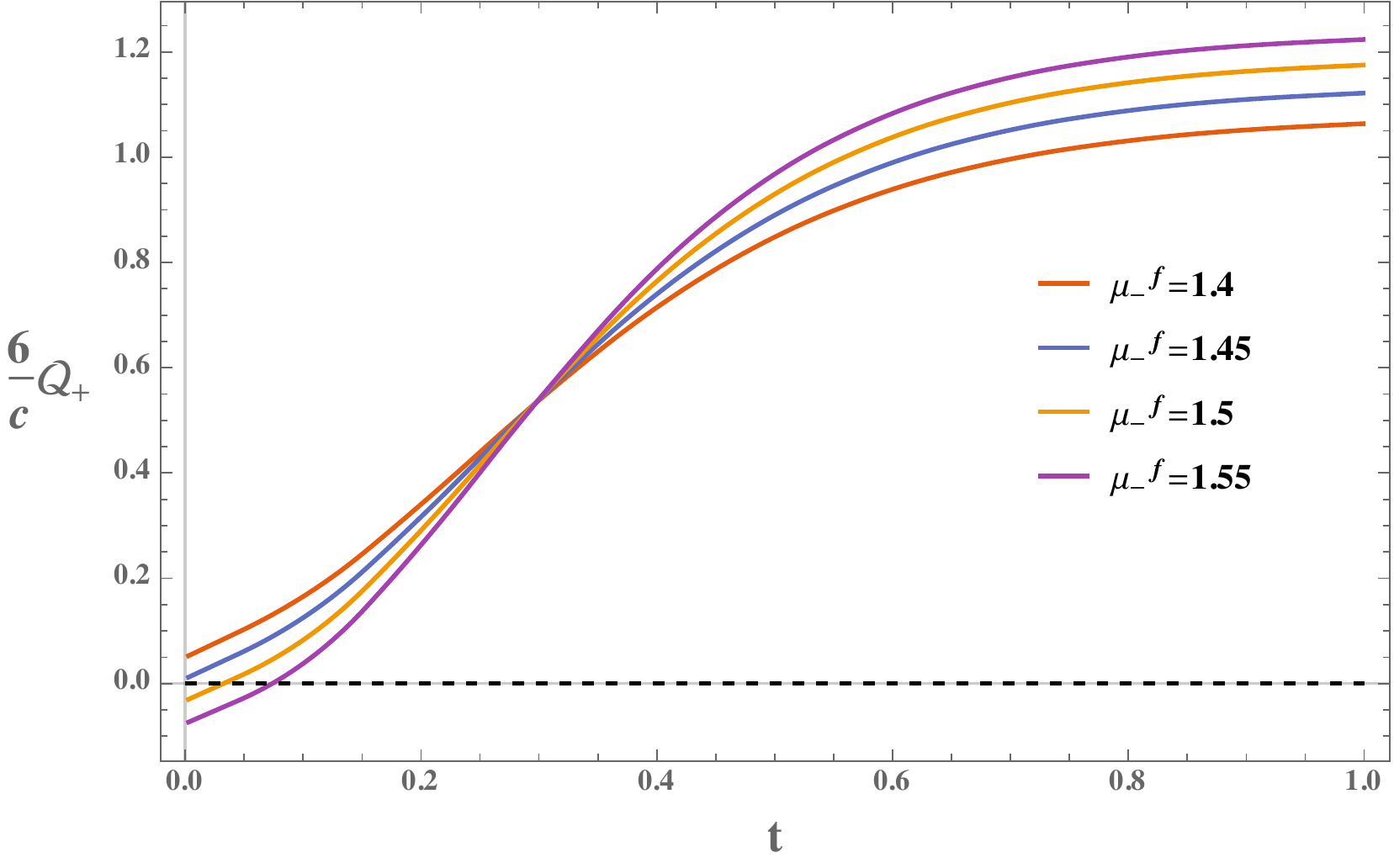}
\caption{$\mathcal{Q}_+$ as a function of time is plotted above for $\mu_\pm^i =1$ and $\mu_+^f=1.2$, and various values of $\mu_i^f$. Note that \eqref{Eq:mufallowed} implies $\mu_-^l \approx 1.08$ and $\mu_-^u \approx 1.46$. We see that  $\mathcal{Q}_+$ increases monotonically and saturates to the value given by Eq. \eqref{Eq:QNECplateau} as $t\rightarrow\infty$ if $\mu_-^l<\mu_-^f < \mu_-^u$. 
}\label{Fig:QNECplargel}
\end{figure}

\begin{figure}[h]
\includegraphics[scale=.35]{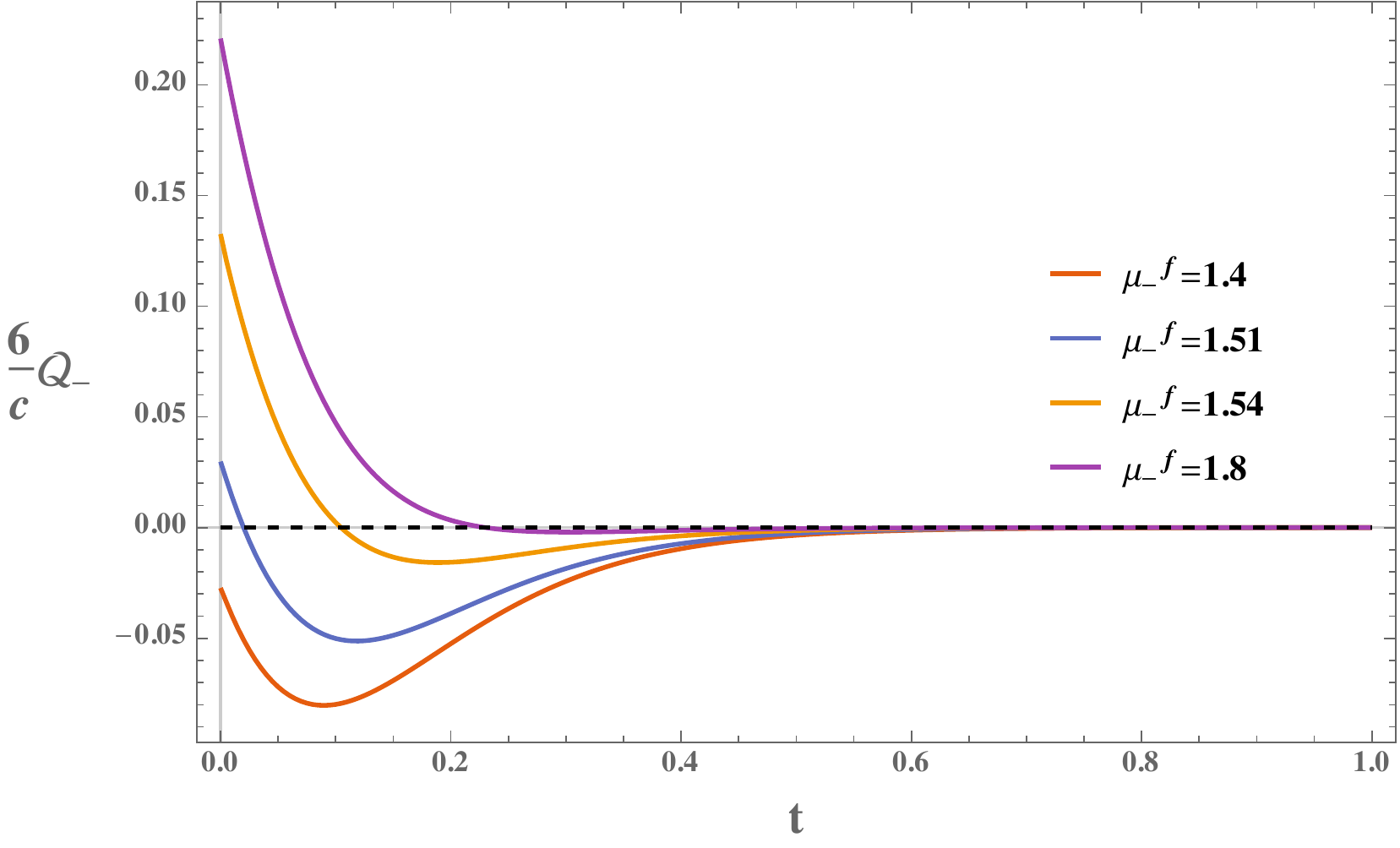}
\caption{$\mathcal{Q}_-$ as a function of time is plotted above for $\mu_\pm^i =1$ and $\mu_+^f=2$, and various values of $\mu_i^f$. According to \eqref{Eq:mufallowed}, $\mu_-^l\approx 1.54$ and $\mu_-^u\approx 2.61$. We see that after a finite time $\mathcal{Q}_-$ becomes negative even if $\mathcal{Q}_-(t=0^+)>0$, i.e. $\mu_-^f > \mu_-^{l}$. However for $\mu_-^{lr}<\mu_-^f< \mu_-^u$, $\mathcal{Q}_-(t) >0$ for $t>0$ and is also a monotonically decreasing function which vanishes at late time. Here $\mu_-^{lr}\approx 1.6$. 
}\label{Fig:QNECmlargel}
\end{figure}

$\mathcal{Q}_+(t)$ increases monotonically in time if $\mathcal{Q}_+(t=0^+) > 0$ and saturates to the value given by Eq. \eqref{Eq:QNECplateau} as $t\rightarrow\infty$ (see Fig.~\ref{Fig:QNECplargel} for plots). It is therefore sufficient to check if $\mathcal{Q}_+(t=0^+) > 0$ to satisfy $\mathcal{Q}_+(t)> 0$ for $t>0$. This is guaranteed if $\mu_-^l < \mu_-^f < \mu_-^{u}$ as illustrated in Fig.~\ref{Fig:QNECplargel}.

$\mathcal{Q}_-(t)$ can be a non-monotonic function even if $\mathcal{Q}_-(t=0^+)>0$. However, there exists $\mu_-^{lr}$ such that $\mu_-^u>\mu_-^{lr}> \mu_-^l$  for which $\mathcal{Q}_-(t)$ is positive for all $t>0$ and is also a monotonically decreasing function if $ \mu_-^{lr}<\mu_-^f$. See Fig.~\ref{Fig:QNECmlargel} for an illustration.  

To find the final allowed quenches, where both $\mathcal{Q}_\pm (t)$ are non-negative for $t\geq0$, it is therefore sufficient to restrict $\mu_-^f$ within the upper and lower bounds set by Eq. \eqref{Eq:mufallowed} for given $\mu_\pm^i$, and then scan upwards by increasing $\mu_-^f$ from the lower end $\mu_-^l$ for fixed $\mu_+^f$ until we find $\mu_-^{lr}$ such that $\mathcal{Q}_-(t)$ stays positive for all $t>0$. This gives us the allowed final states $\mu_-^{lr}(\mu_+^f)\leq\mu_-^f\leq\mu_-^u(\mu_+^f)$ for a given $\mu_\pm^i$ (the white regions shown in Fig.~\ref{Fig:QNEC}). In this allowed region, $\mathcal{Q}_+$ and $\mathcal{Q}_-$ are monotonically increasing and decreasing functions of time respectively (it is easy to check analytically via \eqref{Eq:QNECnonrot} when the final non-rotating state is allowed).
\subsection{The allowed final states for an initial non-rotating state}

\begin{figure}[h]
\includegraphics[scale=.45]{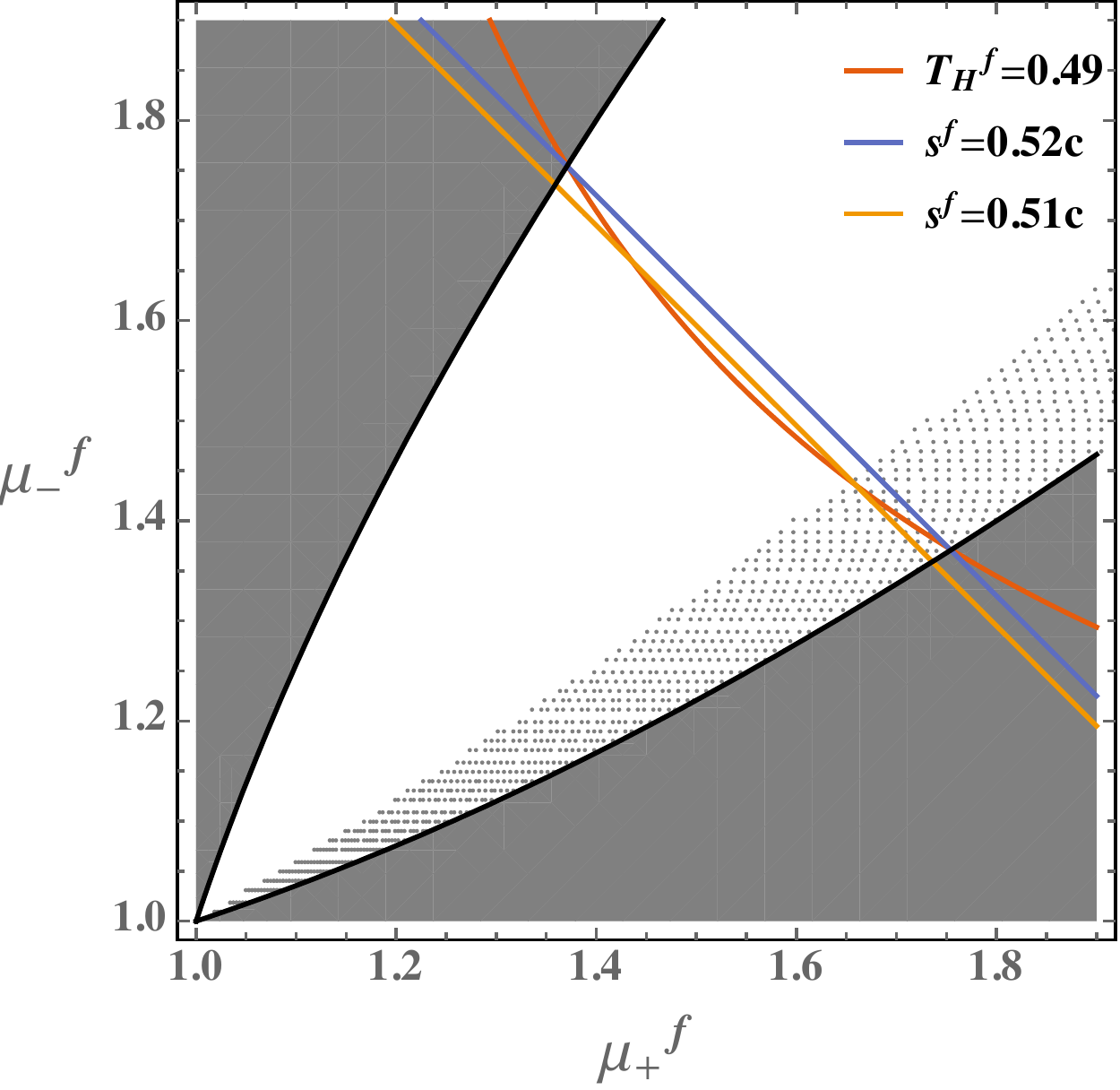}
\caption{The allowed final states corresponding to the initial state $\mu_\pm^i$ is shown in white above. As in Fig. \ref{Fig:QNEC}, the black lines are given by $\mathcal{Q}_\pm \geq 0$ at $t=0$ and the grey dotted region is disallowed by examining $\mathcal{Q}_-$ for $t>0$. For a given final temperature, there is a lower and upper bound on the final entropy density.
}\label{Fig:nonrotallowed}
\end{figure}

The plots for the allowed final states corresponding to an initial non-rotating state has been shown in Fig. \ref{Fig:nonrotallowed}. It provides another illustration of lower and upper bounds on irreversible entropy production for a given final temperature as in Fig. \ref{Fig:QNEC}. 

\subsection{Lower and upper bounds on $D_s$}

\begin{figure}[h]
\includegraphics[scale=.45]{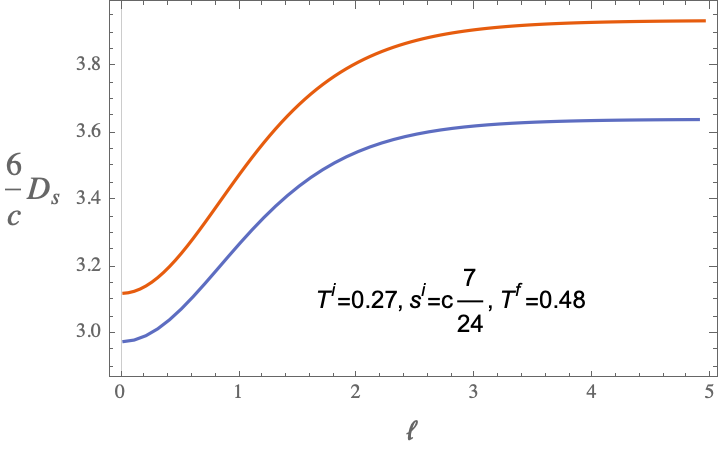}
\caption{ The upper and lower bounds on $D_s$ for a fixed initial state and a final temperature are plotted above as a function of $l$ in red and blue respectively. Both of these bounds increase with $l$ monotonically.
}\label{Fig:diffbound}
\end{figure}

One can convert the restriction $\mu_-^{lr}(\mu_+^f)\leq\mu_-^f\leq\mu_-^u(\mu_+^f)$ set by the QNEC for a fixed initial $\mu_\pm^i$ to a lower and upper bound for the entropy density (temperature) for a fixed final temperature (entropy density) via \eqref{Eq:T-s}. Substituting these in \eqref{Eq:Ds}, one readily obtains a lower and an upper bound on $D_s$ for a fixed initial temperature and entropy density, and a fixed final temperature (as shown in Fig.~\ref{Fig:diffbound}) or for a fixed final entropy density. Both the lower and upper bounds increase monotonically with $l$.
\bibliographystyle{apsrev4-1}
\bibliography{qtquench}

\end{document}